\documentclass[usenatbib,twocolumn]{mn2e}
\usepackage{graphicx}
\usepackage{amssymb}

\usepackage{subfigure}
\newcommand{\beq}{\begin{equation}}
\newcommand{\eeq}{\end{equation}}

\def\apj{ApJ}

\def\beq{\begin{equation}}
\def\ee{\end{equation}}
\def\lsim{\mathrel{\rlap{\lower4pt\hbox{\hskip1pt$\sim$}}
    \raise1pt\hbox{$<$}}}
\def\gsim{\mathrel{\rlap{\lower4pt\hbox{\hskip1pt$\sim$}}
    \raise1pt\hbox{$>$}}}

\def\ts{\times}
\def\lb{\langle}
\def\rb{\rangle}
\def\curl{\nabla {\ts}}

\def\bfv{{\bf v}}

\def\bfb{{\bf b}}

\def\bbB{\overline {\bf B}}

\title[Stellar Activity vs. Rotation  Explained]
 { Explaining the Observed Relation Between Stellar Activity and Rotation }

\author [Blackman and Thomas]
{Eric G. Blackman$^{1,2}$\thanks{E-mail: blackman@pas.rochester.edu}
 \thanks{Simons Fellow; IBM-Einstein Fellow }
and
John H.  Thomas,$^{1,3}$\thanks{E-mail: thomas@me.rochester.edu}
\\ $^{1}$Department of Physics and Astronomy, University of Rochester, Rochester NY, 14627, USA\\
 $^{2}$School of Natural Sciences, Institute for Advanced Study, Princeton NJ, 08540 USA\\
 $^{3}$Department of Mechanical Engineering, University of Rochester, Rochester NY, 14627, USA\ 
 \ }
\begin{document}
\date{}
\pagerange{\pageref{firstpage}--\pageref{lastpage}} \pubyear{}
\maketitle
\label{firstpage}
\begin{abstract}
Observations of late-type main-sequence stars  have  revealed  empirical scalings of coronal activity versus rotation period or
Rossby number $Ro$  (a ratio of  rotation period to convective turnover time) 
which  has hitherto lacked  explanation. For $Ro >> 1$,  the  activity observed as   X-ray to bolometric flux 
varies as  $Ro^{-q}$ with $2\le q \le 3$, whilst $|q| < 0.12$  for $Ro << 1$.  
Here we  explain  the transition between these two regimes and the  power law in the $Ro >> 1$ regime by  constructing
an expression for the coronal luminosity based on dynamo magnetic field generation and magnetic buoyancy.  We  
explain the $Ro<<1$ behavior from  the inference that observed rotation is correlated with  internal differential rotation and argue that once the shear time scale is shorter than the convective turnover time, eddies will be shredded on the shear time 
scale and so the eddy correlation time actually becomes the shear time and the convection time drops out of the 
equations.  We explain the $Ro >> 1$ behavior using a dynamo 
saturation theory based on magnetic helicity buildup and buoyant loss.


\end{abstract}


\section{Introduction}

Observed relations between  coronal activity and rotation period in low-mass stars (Pallavicini et al. 1981; Noyes et al. 1984; Vilhu 1984;  Micela et al. 1985;  Randich 2000; Montesinos et al. 2001; Wright et al. 2011; Vidotto et al. 2014; Reiners et al. 2014) have challenged  theorists.   
A measure of  activity is  the total X-ray luminosity $\mathcal{L}_X$, expressed   as 
\beq
{\mathcal{L}_X\over \mathcal{L}_*} \propto Ro^{-q},
\label{basic}
\eeq  
where $\mathcal{L}_*$ is  the bolometic luminosity,   $Ro$ is the Rossby number,  $Ro \equiv {1/ \Omega \tau_c} ={\tau_r/ 2\pi \tau_c}$, where $\Omega$ 
is the  surface angular velocity, $\tau_r$ is the rotation period, and  $\tau_c$ is the convective turnover time.
For  $Ro>> 0.12$,   the data show that $2\le q \le 3$ , whilst for $Ro < <0.12$, the data show that $|q|<0.2$ (Wright et al. 2011; Reiners et al. 2014).

While $\tau_r$ can be inferred directly from time-series photometry of variability associated  with star spots,
$\tau_c$  is typically  inferred from stellar models 
using the technique of Noyes et al. (1984). In this approach,  $\tau_c= \Lambda h_p/v$  where $h_p$ is  the pressure scale height 
at the base of the convection zone,  $v$  is a convective velocity, and $\Lambda$ is a dimensionless mixing-length parameter (e.g. Shu 1992).  
Stellar model values of $h_p$ and $v$ produce a specific color index such as $B-V$, that can be compared with observations to obtain 
 $\tau_c(B-V)$.   The ill-constrained $\Lambda$ is typically chosen in the range $1<\Lambda < 3$.

Associating  $\mathcal{L}_X$ with  magnetic activity arises from a paradigm in which some fraction of the magnetic field energy 
created within the star by dynamo action rises buoyantly through the star and ultimately
 converts some of its energy into accelerated particles that  radiate as coronal X-rays  
 (e.g Schrijver  \&  Zwaan 2000). 
The  connection between X-ray activity, magnetic field generation, and  rotation and 
differential rotation for dynamos (e.g. Moffatt 1978; Parker 1979 Krause and R\"adler 1980), has  led to the notion that increased 
activity has something to do with efficiency of dynamo action (Noyes et al. 1984; Montesinos et al. 2001; Wright et al. 2011) but  
connecting this to a  theoretical explanation of Eq. (\ref{basic}) has  been lacking.

Previous efforts to  explain Eq. (\ref{basic}) 
have focused on the dimensionless dynamo number.
In Sec. 2 we show that such approaches  used in previous work 
are invalid for $Ro<1$. After revising these estimates, we  then argue that the more conceptually relevant quantity for connecting X-ray activity 
with dynamo action is in fact  the saturated field strength before magnetic buoyancy ensues, which we derive. The saturated value we derive emerges with a scaling consistent with that shown to match a 
wide range of simulations (Christensen 2009).
In Sec. 3 we use our results of section 2 to derive an expression for $\mathcal{L}_X/\mathcal{L}_*$.
 We conclude in Sec. 4.

\section{Rethinking Key Dynamo Quantities}

How  and where the dynamo operates in solar-like
stars remains an open question (see Charbonneau 2014). Interface dynamos,  in which 
the shear layer of a tachocline beneath the convection zone dominates the  toroidal field amplification by 
differential rotation ($\Omega$- effect) while the convection zone above provides the  helical $\alpha$ 
effect, have been proposed in order to avoid the problem that field strengths greater than a few 100 Gauss 
might rise too quickly through the convection zone and thus could not easily be generated therein (e.g. Deluca \& Gilman 1986; Parker 1993; 
Thomas et al. 1995; Markiel \& Thomas 1999). 
 The low latitudes of flux emergence and the small tilt angles 
of loops anchored at sunspots indicate local field strengths of order $10^5$ Gauss is needed to avoid too much deflection 
of rising flux tubes by the Coriolis force. Such strong fields are most easily anchored in the tachocline.
(The field strength in local structures is much larger than that of the  spatially averaged mean field.)

However,  Brandenburg (2005) emphasizes that there remain plausible arguments   that the dynamo could  be more distributed in the convection zone,
  akin to  original models (Parker 1955; Moffatt 1978; Parker 1979 Krause \& R\"adler 1980). Downward turbulent pumping  can substantially reduce the rate of buoyant rise of flux tubes
(e.g. Hurlburt et al. 1984; Tobias et al. 2001; Thomas et al. 2002; Brummell et al. 2008), perhaps obviating one of the motivations for the interface 
dynamo. Also, regions of strong shear near the tachocline are located at high latitudes,
not low latitudes where sunspots appear. In contrast, the near-surface shear layers are closer to latitudes where the 
sunspots appear.  In addition, surface layers show rotational variations on the 
solar cycle time scale. Finally, the fact that fully convective M stars have dynamos and activity  shows that an interface dynamo 
is not generally needed, although this does not preclude its existence in higher-mass stars.

Here we revisit  previous parameter scalings for both interface and distributed dynamos that are valid for $Ro>>1$ 
but have been unwittingly used when $Ro << 1$ .   We revise them by taking into account the fact that the eddy correlation time 
is not the convective turnover time for $Ro << 1$.    We also estimate  the saturated magnetic field strength, which we argue 
to be most important for the  activity-Rossby number relation we derived in section 3.

\subsection{ Dynamo Number  for  $Ro << 1$ and $Ro >> 1$ }

The  spherical $\alpha-\Omega$ mean field dynamo equations can be written 
(Durney \& Robinson 1982; Thomas et al. 1995)
\beq
{\partial A_\phi \over \partial t}=\alpha B_\phi+\beta_1 \left(\nabla^2 - {1\over r^2 sin^2 \theta}\right) A_\phi
\label{d1}
\eeq
and
\beq
{\partial B_\phi \over \partial t}=r sin \theta ({\bf B}_p\cdot \nabla )\Omega +\beta \nabla^2 B_\phi
\label{d2}
\eeq
where the large-scale magnetic field is ${\bf B}=B_\phi{\hat \phi} + \curl  (A_\phi{\hat \phi})$, $B_\phi$  and $A_\phi$ are  the 
large-scale  toroidal magnetic field and vector potential components,  ${\bf B}_p=\curl  (A_\phi{\hat \phi})$  is the poloidal  
magnetic field component, and $(r,\theta,\phi)$ are spherical coordinates. The angular velocity $\Omega=\Omega (r,\theta)$
in general. In  Eqs. (\ref{d1}) and (\ref{d2}), $\alpha$,  $\beta_1$ and $\beta$ are respectively pseudoscalar helicity 
and scalar diffusion transport coefficients (discussed later)  that incorporate turbulent correlations of  in the electromotive force (EMF) 
$\lb\bfv \times \bfb \rb= \alpha \bbB -\beta \curl \bbB$.  

We remove the $r$ dependence in  Eqs. (\ref{d1}) and (\ref{d2})  by  assuming that the poloidal variation dominates the radial variation  (Yoshimura 1975;  Durney \& Robinson 1982) and  write  
$ {A_\phi} sin \theta = A(t) e^{i\theta k  r_c}$, and ${B_\phi}  = B(t) e^{i\theta k r_c}$, where $r_c$ is a radius 
in the convection zone nearest to where the shear is strongest (i.e. the base for the sun)
 and $k$ is the wavenumber associated with the radius of curvature  
for quantities dependent on $\theta$. With a  buoyancy loss term  added to to Eq. (\ref{d2}), 
Eqs. (\ref{d1}) and (\ref{d2})  then become (Durney \& Robinson 1982)
\beq
{\partial A\over \partial t}=\alpha B-\beta_1 k^2 A
\label{d3}
\eeq
and
\beq
{\partial B \over \partial t}=i k r_c \Delta \Omega {A\over L}-\beta k^2 B- {u_bB\over L},
\label{d4}
\eeq
where $L$ is the thickness of the shear layer of differential rotation $\Delta\Omega$, 
 and $u_b$ is a buoyancy speed.
Eqs. (\ref{d3}) and (\ref{d4}) can be applied  to a distributed dynamo or an interface dynamo.
For a  distributed dynamo $L$ is the width of the convection zone. For an  interface dynamo, Eqs. (\ref{d3}) and (\ref{d4}) 
provide a  1-D approximation where each layer is separately assumed to have fields that vary slowly in radius 
(e.g. Thomas et al. 1995), and where $L$ is the thickness of the shear layer just beneath the convection zone 
and $L_1$ is the thickness of the layer  above $r=r_c$ where the $\alpha$ effect operates.

To proceed, we use a standard substitution $A(t)= A_0e^{-i\omega t}$  and $B(t)= B_0e^{-i\omega t}$ 
with $\omega= \omega_R +i \omega_I$, where $\omega_R$ and $\omega_I$ are real. We assume that  
buoyancy  kicks in when the field has reached a value beyond that attained in the early-time kinematic 
growth phase, so we here estimate the growth condition without the last term in Eq. (\ref{d4}).
Eqs. (\ref{d3}) and (\ref{d4}) have growing solutions when the absolute value of the product of the growth  coefficients of the two equations divided by the product of the decay coefficients)
   \beq
 N_D={\alpha_0  r_c  \Delta \Omega \over L\beta_1\beta   k^3}, 
 \label{nd1}
 \eeq
exceeds unity. This $N_D$ is the dynamo number.

We now specify explicit expressions for  $\alpha_0$, $\beta_1$, and $\beta$.
An estimate of $\alpha_0$ that incorporates the Coriolis force  is (Durney \& Robinson 1982):
\beq
\alpha_0={\tau_{ed}\over 3} \lb\bfv\cdot\curl\bfv \rb \sim {q_\alpha\over 6}\tau_{ed}^2 {\Omega v^2\over r_c} cos \theta,
\label{4}
\eeq
where $\bf v$ is the turbulent convection velocity of magnitude $v$ and 
$\tau_{ed}$ is the  eddy correlation time (assumed short enough to replace a time-correlation integral e.g Moffat 1978, although see  Blackman \& Field (2002)  for a  different interpretation of $\tau_{ed}$.) in the convection zone. Note that  $\tau_{ed}$ need not equal  $\tau_c$ since the latter is inferred from the pressure scale height and 
convective velocity (Noyes et al. 1984).
The constant $q_\alpha = [{\Omega(r_c)/ \Omega}] [{k^2_{v,\theta}/(k^2_{v,\phi}+k^2_{v,\theta})}]$
 accounts for both the factor by which the angular velocity at the base of the convection zone differs from that  at the surface, and  anisotropy of  turbulent  wave vectors.
We take  
\beq
\beta_1 = {1\over 3 }v^2 \tau_{ed}=  q_\beta \beta , 
\label{4b}
\eeq
where  $q_\beta\ge 1$ is a constant.

 Eqs. (\ref{4}) and (\ref{4b}) apply only  for $Ro>>1$,  for which $\tau_{ed}=\tau_c$.
Their invalidity for $Ro<<1$ is evident from Eq. (\ref{4}):  when $\tau_{ed}\Omega >>1$, 
the magnitude of $\alpha_0$, which has dimensions of velocity, could otherwise exceed $v$ (Robinson \& Durney 1982). 
Only $|\alpha_0|< v$ can be physical, since only a fraction  $\le 1$ of the velocity is helical, and the gradient scales 
entering $\alpha_0$ can be no smaller than that of the dominant  eddy scale.  This  restriction on the regime of validity  
has not been taken into account previously in studies where $N_D$ is presented  in the context of activity versus 
Rossby number relations (e.g. Montesinos et al. 2001; Wright et al. 2011).

However, we now extend the regime of validity  by a physically motivated redefinition of $\tau_{ed}$. 
A strong surface rotation  is  plausibly also indicative of  strong differential rotation within the star and 
  if a  convective  eddy is shredded by shear  on a time scale $\tau_s<\tau_c$,  then
 the shorter shear time scale $\tau_s$ becomes the  relevant eddy correlation lifetime such that  $\tau_{ed}\sim \tau_s$.  
 We assume  $\tau_s = s \tau_r$ where $s$ is a constant that accounts for  differential rotation.
 For  $2\pi s Ro={s \tau_r\over \tau_{c}}<<1$  
we have  $\tau_{ed}\sim s\tau_r << \tau_c$ so $Ro$ should drop out of the equations.
The  transition will occur approximately at  $2\pi Ro \sim 1/s$.
As discussed further later,  we interpret the  observational transition 
where the activity becomes independent of $Ro$ as exactly the transition to this shear dominated regime.
An empirical transition at  $2\pi Ro \sim 0.12$ (Wright et al. 2011) would imply $s\simeq 8.3$.

To  capture both  regimes $Ro>>1$ and $Ro<<1$ based on the physical argument just presented, we  write 
\beq
\tau_{ed}= {s\tau_r\over 1+ 2\pi s Ro}.
\label{taued}
\eeq
The dynamo coefficients (and thus  the EMF) decrease with shear,
consistent with simulations of Cattaneo \& Tobias (2014).
Using $\tau_r={2\pi/ \Omega}$ with Eq. (\ref{4}) and (\ref{4b}) with   Eq. (\ref{taued}) in Eq. (\ref{nd1}),
along with $\Delta \Omega= \Omega/s$ and $kr_c \sim1$,
we obtain
 \beq
 N_D \sim { 3  q_\alpha q_\beta  \Omega^2 r_c^3  cos \theta_s  \over 2s L  v^2} = 
  {3  q_\alpha q_\beta r_c^3 cos \theta_s  \over 2 sL v^2\tau_{ed}^2} 
  \left({2\pi s\over 1+2\pi s Ro}\right)^2,
\label{nd2}
 \eeq
 where $\theta_s$ is a fiducial value for colatitude associated with $s=s(\theta_s)$. 
 For  $Ro>>1$, $\tau_{ed}\sim \tau_c$ from Eq. (\ref{taued}), and 
 we write $v\tau_{ed} \simeq v\tau_c\sim  h_p$, where $h_p$  is a pressure scale height
 satisfying     $L_1 \sim \xi h_p$, where the constant $\xi \sim\ $few.
Thus 
 \beq
 N_D (Ro >>1) \sim
  {3 q_\alpha q_\beta cos \theta_s \xi^2 r_c^3 \over 2 s  LL_1 ^2}  Ro^{-2},
\label{nd2large}
 \eeq
highlighting the $Ro^{-2}$ scaling as in Montesinos et al. (2001) but with different coefficients 
in part because we have used a more general formula for $\alpha_0$.
Note that Eq. (\ref{nd2large}) applies only for  $Ro>>1$ as discussed above. 
If we  assume a distributed dynamo, for which  $L\sim L_1$ and $q_\beta =1$, 
then 
 \beq
 N_D (Ro >>1, dist) \sim 
  {3   q_\alpha cos \theta_s \xi^2 r_c^3 \over 2 s  L^3}  Ro^{-2}.
\label{nd2large2}
 \eeq
 
For  $Ro << 1$, we have $\tau_{ed} << \tau_c$ and so $\tau_{ed}\propto\tau_r$  in Eq. (\ref{taued}). In this regime  
$\Omega/v  \sim 2\pi s / l_{ed} $, where $l_{ed}$ is the eddy scale. Eq. (\ref{taued}) and (\ref{nd2})  then imply that 
 \beq
 N_D (Ro<<1) \sim { 6 \pi^2  s q_\alpha q_\beta  r_c^3 cos \theta_s  \over  L  l_{ed}^2} .
\label{nd2small}
 \eeq
If we further assume a distributed dynamo so that $L\sim L_1$, and $q_\beta =1$
then 
 \beq
 N_D (Ro<<1, dist) \sim { 6 \pi^2  s q_\alpha  r_c^2 cos \theta_s \over  L l_{ed}^2} .
\label{nd2small2}
 \eeq

Previous discussions linking activity to dynamos have focused on the  $Ro$  dependence of $N_D$ 
using the $Ro >> 1$ formulae  (Montesinos et al. 2001; Wright et al. 2011) but without making a specific
 theoretical connection to  coronal luminosity.  We argue in Sec. 4 that  $N_D$ 
is not the most  important quantity for  predicting the activity-$Ro$ relation.

\subsection{Saturated Field Strength: Estimate and Role}

Although $N_D$ determines the kinematic cycle period and growth rate 
 it does not  determine the nonlinear cycle period (e.g. Tobias 1998),  
 nor  the saturated magnetic field strength.
  The saturated  dynamo field strength is in fact commonly imposed by hand (e.g. Markiel \& Thomas 1999; 
Montesinos et al. 2001; Charbonneau 2014). But the saturated field strength is important for determining how much 
magnetic energy is delivered to the corona and thus the X-ray luminosity averaged over a cycle period.

Recent work has progressed toward a saturation theory that agrees with simulations of simple  helical  dynamos 
when 20th-century textbook mean field  theory is  augmented to include  a tracking of the evolution of magnetic helicity  
(for reviews see Brandenburg \& Subramanian 2005; Blackman 2014).  
When the predominant kinematic driver is kinetic helicity, a key ingredient is that the dynamo $\alpha$  is best represented as the difference  $\alpha_0-\alpha_M$, where $\alpha_M=\lb\bfb\cdot \curl \bfb\rb\tau_{ed}$,  
proportional to the  current helicity density of magnetic fluctuations. This form  emerged from the spectral  approach 
of Pouquet et al. (1976) and  from a simpler two-scale mean-field dynamo approach  (Blackman \& Field 2002; 
Field \& Blackman 2002). In the Coloumb gauge, $\alpha_M$ is proportional to the magnetic helicity density of fluctuations,   
a result that is also approximately true in an arbitrary gauge when the  fluctuation scale is much less than the averaging scale  
(Subramanian \& Brandenburg 2006).   

Saturation in the stellar context might proceed as follows (Blackman \& Brandenburg 2003):  kinetic helicity initially drives 
the large-scale helical magnetic  field growth, which, to conserve magnetic helicity for large $R_M$, builds up small-scale 
magnetic helicity of the opposite sign. This grows $\alpha_M$ to offset $\alpha_0$. To lowest order in 
${ l_{ed}\over  L_1}$,  the greatest strength the large-scale helical field can attain before catastrophically slowing the 
cycle period is estimated by setting  $\alpha_0-\alpha_M\simeq 0$ and using magnetic helicity conservation to connect 
$\alpha_M$ to the large scale helical field.  The toroidal field is further amplified non-helically  by differential rotation 
above the strength of the mean poloidal field to a value which is limited by magnetic buoyancy. We assume that downward 
turbulent pumping (e.g. Tobias et al 2001) hampers buoyant loss only above some threshold field strength (Weber et al. 2011; Mitra et al. 2014) such that $\alpha_M$ can  approach the value  $\alpha_0$.
A sustained dynamo is then maintained with large-field amplification is balanced by  buoyant loss, coupled to a beneficial 
loss of small scale helicity:  for dynamos with shear, small scale helicity fluxes  seem to be essential not only for sustaining 
a fast cycle period but also to avoid catastrophic field decay (e.g. Brandenburg \& Sandin 2004; Sur et al. 2007).  \footnote{Magnetic buoyancy  could initially source the EMF  instead of kinetic  helicity. Upon saturation from small scale twist,   buoyancy would  still eventually act as a loss mechanism.  The connection to flux transport dynamos  is   beyond our present scope but Karak et al. (2014)  present a challenge  that such dynamos predict cycle period-$Ro$ trends opposite to both those observed.}

The saturation strength of the large-scale poloidal field, based on the aforementioned circumstance
of $\alpha_M \sim \alpha_0$ is  (Blackman \&  Field 2002; Blackman \& Brandenburg 2003)  
\beq
B_p^2 \sim  8\pi{l_{ed}\over L_1} f_h \rho  v^2, 
\label{bpol}
\eeq 
where $f_h$ is the fractional magnetic helicity when the initial driver is kinetic helicity. For our present purpose, 
$f_h= {l_{ed}\lb \bfv \cdot \curl \bfv\rb| / v^2} = {q_\alpha cos \theta_s \over 6}{s\over 1+ 2\pi s Ro}{l_{ed}\over r_c} $,
where the latter expression follows from Eqs. (\ref{4}) and  (\ref{taued}).

The toroidal field is linearly amplified by shear above this value during a buoyant loss time 
$\tau_b\sim L_1/u_b$, where $u_b$ is a typical buoyancy speed for those structures that escape.
This gives, 
\beq
{B^2\over 8\pi} \simeq  {B_p ^2\over 8\pi}(\Omega \tau_b/s)^2,
\label{16}
\eeq
where the latter similarity assumes  $B_\phi > B_p$ which is valid long as $\Omega \tau_b /s>1$. If we further assume that 
$u_b\simeq {B_\phi^2/(12 \pi \rho v)}$ from calculations  of buoyant flux 
 rise (Parker 1979; Moreno-Insertis 1986; Vishniac 1995; Weber et al. 2011), 
 then $\tau_b\simeq{ 12\pi L_1  \rho v/ B_\phi^2}$. Using these in 
(\ref{16}) and solving for $B^2\sim B_\phi^2$   gives 
\beq
\begin{array}{r}
{B^2\over 8\pi}\sim   
\left({12 \rho B_p \Omega v L_1\over 8^{3/2} s{\sqrt \pi}}\right)^{2\over 3}
 \sim  {(3\pi)^{2\over 3}} \left({sL_1q_\alpha cos\theta\over 6r_c}\right)^{1\over 3}
 {\rho v^2\over (1+2\pi s R_o)}
\label{bsat}
\end{array}
\eeq 
where we have used the above expression for $f_h$, Eq. (\ref{bpol}),   $\tau_{ed}\Omega = {2\pi l_{ed}\over v \tau_r}$ and
Eq. (\ref{taued}). 
Eq. (\ref{bsat}) leads self-consistently to $B_\phi> B_p$ as long as $\Omega v L_1/s > {B_p^2 /(8\pi \rho) }$, which
is satisfied even for slow rotators like the sun since for $s=5$, $\Omega/s \sim 4 \times 10^{-7}$/s,  $v\sim 4000$cm/s, $L_1 \sim 
10^{10}$cm, and ${B_p^2/ 8\pi} < \rho v^2$ from Eq. (\ref{bpol}). 
Eq. (\ref{bsat}) also agrees with the   $Ro<<1$  scaling of Christensen et al. (2009) which matches planetary and stellar dynamos simulations since  for $Ro<<1$,  $Ro$ drops out 
and $B^2$ becomes independent of $\Omega$, and  $B^2\propto \rho^{1/3} ({\rho v^3})^{2/3}$.


\section{X-ray Activity and Rossby Number }

Overall stellar activity can be gauged by $\mathcal{L}_X/\mathcal{L}_*$, where  $\mathcal{L}_X$ is
assumed to result from dissipation of  dynamo-produced magnetic fields that rise  into the corona.     
Each single observation of $\mathcal{L}_X$ probes a time scale short compared to the cycle period and hence can be thought of 
as taken from an ensemble of luminosities from a distribution over a cycle period. 
The  solar X-ray luminosity varies by s little more than an order of magnitude over 
a solar cycle (Peres et al. 2000), so  even a cycle averaged estimate is meaningful. The saturated value of the field  is then 
 more important than $N_D$.

We estimate the coronal X-ray luminosity as the product of the magnetic energy  flux that results from the dynamo field 
production and buoyant rise into the corona, assumed to be averaged over a dynamo cycle period.
A significant fraction (but not all) of the magnetic energy is kept from rising by the action of downward turbulent pumping  
(e.g. Tobias et al. 2001; Thomas et al. 2001; Brummell et al. 2008).   Focusing on the fraction that rises, we then estimate the X-ray luminosity  as
\beq
\begin{array}{r}
\mathcal{L}_{X}\simeq  \mathcal{L}_{mag} \simeq  {B^2 u_b\over 8 \pi} \Theta r_c^2
= {2\over 3}\left({B^2 \over 8 \pi}\right)^2{ \Theta r_c^2\over \rho v}
\end{array}
\label{lmagsimple1}
\eeq
  where $B\sim {B}_\phi$ and $\Theta$ is the solid angle through which the field rises, and we have used the expression for $u_b$   below Eq. (\ref{16}).
  \footnote{ In Eq. (\ref{lmagsimple1}) ${\mathcal L}_x\propto B^4\propto B_p^{4/3}$, is not inconsistent with  observed relations  $L_x \propto \Phi^m$ $(1<m<2)$, (Pevstov et al 2003; Vidotto et al. 2014) where $\Phi$ is the poloidal flux.}
Using  Eq. (\ref{bsat}) in Eq. (\ref{lmagsimple1})
then gives
 \beq
\begin{array}{r}
{\mathcal{L}_{X} \over \mathcal{L}_* }\simeq
 \left({L_1\over r_c}\right)^{2/3}\left({s^{1/3}\over 1+2\pi sRo}\right)^{2} \Theta \left({3\pi\over 8 }\right)^{1\over 3}\left({q_\alpha cos \theta_s \over 6}\right)^{2/3}.
\end{array}
\label{lmagsimple3}
\eeq
where   we have used
 $
\mathcal{L}_*\simeq  \mathcal{L}_{c} 
  \simeq   4\pi r_c^2\rho {v^3},
$
 the luminosity  associated with the convective heat flux 
through the convection zone (e.g. Shu  1992).
For the sun $2\pi Ro\sim 2$ and $L_1\sim 2r_c/5$. 
We posit that $\Theta/4\pi $  is proportional to the  areal fraction through which  the strongest buoyant fields   penetrate. This is likely related  to the areal fraction of sunspots $a_{spt}$.
For the sun  $a_{spt} \lsim 0.005$   (Solanki and Unruh 2004).  For  very  active stars $a_{spt} \ge 0.01$ (O'Neal et al 2004). 
Fig. 1 shows the result of Eq. (\ref{lmagsimple3}) for  $q_\alpha cos \theta_s \sim 0.1$ and   $\Theta=\Theta_0[({\mathcal L}_X/ {\mathcal L}_*)/ 6.6\times 10^{-7}]^{\lambda}$
for two cases of $\lambda=0$ and one case of $\lambda=1/3$, normalized by the average solar value (Peres et al. 2000).

From Eq. (\ref{lmagsimple3}) for  $Ro>>1$,  the $\lambda=0$ cases gives
$q\sim 2$ and  the $\lambda= 1/3$ case gives $q=3$.  Larger $\lambda$ would make $q>3$, whereas the range  $2 \le q\le 3$ is suggested by observations  (Wright et al. 2011).   A curve with $\lambda> 0$ can accommodate  the higher observed saturation values of $\mathcal{L}_X$, compared to the curves of $\lambda=0$, while still matching the sun.  

Most importantly, note that the expression for $\mathcal{L}_x/\mathcal{L}_*$ in Eq. (\ref{lmagsimple3}) becomes independent 
of $Ro$ for $Ro <<1$, regardless 
of the specific behavior in the  $Ro >> 1$ regime. 
\begin{figure}
\centering \includegraphics[width=\columnwidth]{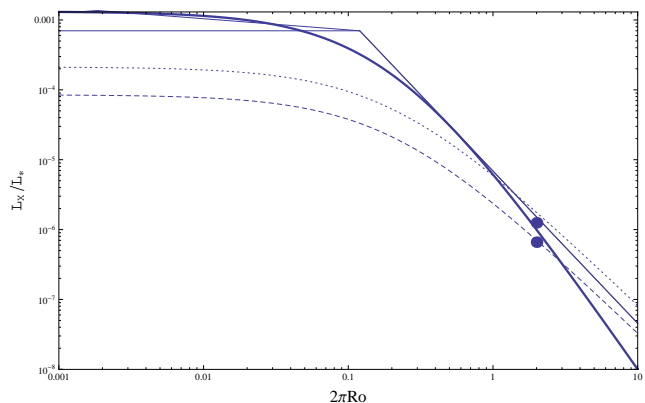}
 \caption{${\mathcal{L}_{X}/\mathcal{L}_{*}}$ from Eq. (\ref{lmagsimple3}). Dashed curve: 
 $\Theta_0= 4.3\times 10^{-4}$;  Dotted curve:  $\Theta_0=1.1\times 10^{-3}$,  both 
 for  $\lambda=0$  and thus $q=2$ (see text);   Solid curve: $\Theta=  2.6\times 10^{-3}[(\mathcal{L}_{X}/\mathcal{L}_*)/(\mathcal{L}_{X}/6.6\times 10^{-7}]^{1/3}$, corresponding to  $q = 3$. All cases use $q_\alpha cos \theta_s=0.14$.
 The  dots correspond to the solar average (lower) and  maximum (upper)  (Peres et al. 2000) and the straight  line to the right of $2\pi Ro =0.12$ is  the Wright et al. (2011) data fit.  
  Straight lines to the left of this point correspond to   $q=0$ (Wright et al. 2011) and
 $q=0.16$ (Reiners et al. 2014)  data fits.}

\end{figure}

\section{Conclusion}

Using  physical arguments, we have developed a  relationship between $\mathcal{L}_X/\mathcal{L}_*$ and  $Ro$.  The result accounts for both a transition to $Ro$ quasi-independence at low $Ro$ and a strong inverse dependence at large $Ro$, in general agreement with observations. 
Our   result that the
 predicted transition toward $Ro$ independence at low $Ro$-- 
 is independent of the specific  dependence on $Ro$ for $Ro >> 1$.  
 Our emergent saturated field strength s for $Ro<<1$ also agrees with the scalings of Christensen (2009), shown to match  a range of planetary and stellar dynamo simulations.

Previous attempts to explain the activity--$Ro$ number relation have focused on the possible role 
of the dynamo number but  the expressions commonly used are invalid for  $Ro << 1$ limit because the  convection turnover time  is no longer a good approximation for the turbulent correlation time.  When eddies are sheared faster than convection can overturn them,  the shear time should replace the convection time 
when estimating correlation times. We have accounted for this using Eq (\ref{taued}), which reduces to  $\tau_c$ for $Ro >> 1$ and to $\tau_r$ for $Ro << 1$. This prescription is  widely applicable.  

More fundamentally,  the dynamo number is insufficient for capturing activity because it does not determine the saturated magnetic field strength. We estimated the 
latter using a saturation theory rooted in magnetic helicity evolution, combined with a loss of magnetic field by magnetic buoyancy. 
The associated magnetic flux provides the source of the  X-ray luminosity and, when combined with the generalized $\tau_{ed}$ 
just described, culminates in Eq.(\ref{lmagsimple3}).

Opportunities for further work abound.  Observational  constraints on $\lambda$ and on the connection between rotation and internal differential rotation would be desirable in testing Eq.(\ref{lmagsimple3}).




\section*{Acknowledgments} 
 We thank M.  Ag\"ueros and E. Mamajek  for discussions. EB acknowledges  grant support from 
 NSF-AST-1109285, HST-AR-13916.002, a Simons  Fellowship, and the IBM-Einstein Fellowship Fund at IAS.

\bibliographystyle{mn2e}

\end{document}